\documentclass{svproc}
\usepackage{url}
\usepackage{graphicx,color,amssymb,amsmath}

\newcommand{\be}{\begin{eqnarray}&&}
\newcommand{\ee}{\end{eqnarray}}
\newcommand{\nonu}{ \nonumber \\ &&}
\def\psla{\slash \! \! \!}

\begin{document}
\mainmatter              
\title{Few-body systems in Minkowski space: the  Bethe-Salpeter Equation 
challenge}
\titlerunning{FB Bethe-Salpeter Eq.}  
%
\author{Giovanni Salm\`e}
\authorrunning{Giovanni Salm\`e} 
%

\institute{INFN, Sezione di Roma,  P.le A. Moro 2,
I-00185 Rome, Italy
}

\maketitle              

\begin{abstract}
Solving the homogeneous Bethe-Salpeter equation 
directly in Minkowski space is  becoming a very  alive field, since,  in recent years,  a 
new 
 approach has been introduced, and the reachable results can be potentially useful in various areas
 of research, as soon as the relativistic description of few-body bound systems
is
  relevant. A brief review of the status of the novel approach,
  which benefits from the consistent efforts of different groups, 
  will be presented,  together with some recent results
\end{abstract}
\section{Introduction}
A fully covariant and  non perturbative
description of a bound system, with and without spin degrees of freedom (dofs),
represents a Holy Grail in different areas of research, as soon as 
  relativistic effects become urgent for obtaining a reliable 
 control  of the dynamical evolution of the state. Indeed, a  bound system
 has a unavoidably non perturbative nature, and if it is necessary a
 relativistic description, one keenly quests a suitable
  approach with
  the above mentioned features
 for tackling the issue. In view of this,
the  Bethe-Salpeter equation (BSE) in Minkowski space can play a role, 
since it was proposed by Salpeter and Bethe \cite{BS51} within a  fully field-theoretical and non perturbative
framework (see Ref. \cite{Nakarev} for a first review). In particular, the capability to offer a non perturbative 
description has to be ascribed to its-own structure, since the BSE for 
bound states is a homogeneous integral equation, and hence it is able to
 take into account all the infinite exchanges, needed to reproduce the bound-state pole (in the relevant
 Green's function).
Given such an attractive feature, the BSE has been actively studied, but the 
 presence of many  challenges related to, e.g., the need of kernels beyond the 
 ladder one (see the next Section),  the presence of self-energy and 
 vertex corrections and,  last but not least, the non trivial  analytic 
 structures of the BS amplitude (i.e. the solution of the BSE), has urged 
 the introduction of wise mathematical
 trick, like the Wick rotation \cite{wick1954properties} and/or sharp approximations.  
 As is well-known,  very popular
approaches to the BS formalism are based on the rotation to imaginary 
relative energy (see, e.g.
Ref. \cite {Tjon80} for an application to the deuteron)  or on  productive  
3D-reductions of
 the BSE (see e.g., Ref. \cite{Lucha05} for some introductory details), and they are devised for
    circumventing the  issue of the cumbersome
 analytic structure in Minkowski momentum-space, 
 related to  the role of
 the relative energy (as well as the meaning of the relative time).  
 There exist also covariant approximation of the BSE (see, e.g.,  Ref. \cite{Gross:2017ipu} 
 illustrating the covariant spectator model approach and Ref. \cite{Bondarenko}
  for the approach based on separable kernels), where some simplifying 
  assumptions are introduced but still keeping safe the covariant nature 
  of BSE.  Till nineties, only one case 
  was known to be  exactly
solved in Minkowski momentum-space, though with the interaction kernel in ladder
approximation, namely  
the Wick-Cutkosky model \cite{wick1954properties,cutkosky1954solutions}, which is 
composed by two massive scalars interacting through   a  {\em massless} scalar.
In 1997, a new approach, based on  i) the so-called Nakanishi Integral Representation (NIR) 
of the BS amplitude\cite{nak63} and ii)  its uniqueness,  was introduced in order to 
  successfully solve the ladder BSE governing a two-scalar system 
interacting through a {\em massive} scalar \cite{Kusaka}. In particular,   
the emphasis of the work was on the calculations  of both  coupling constants 
needed for binding 
the system with assigned masses, and  the so-called 
Nakanishi weight functions (see Sect. \ref{sec_nir}), rather than
on the evaluation of the momentum distributions, so important in the phenomenological 
investigation of the inner dynamics of the bound systems. Nonetheless,
Ref. \cite{Kusaka} opened a new and attractive path, though
the approach  involved a lengthy algebraic manipulations. A substantial improvement was reached
through  the clever introduction \cite{CK2006} of the light-front
(LF)
framework (see e.g., Refs  \cite{Brod_rev,CK_rev} for detailed reviews). Indeed,
 by adopting  LF variables, $x^\pm=x^0\pm x^3$ 
and
${\bf x}_\perp\equiv \{x^1,x^2\}$, it is possible 
to simplify the treatment of the needed analytical integrations, extending
 the realm of application  even including  bound system with spin dofs \cite{CK2006b,dFSV1,dFSV2}, and,
 more important, a probabilistic framework can be established.
 
In Sects. \ref{sec_bse} and \ref{sec_nir} some snapshots on the Bethe-Salpeter equation  and
the Nakanishi integral representation of the BS amplitude will be presented, while Sect.
\ref{sec_lf} is  devoted to the LF framework. Finally 
Sects. \ref{sec_ris} and \ref{sec_con} contain some numerical results and 
perspectives, respectively
\section{The Bethe-Salpeter equation in a nutshell}
\label{sec_bse}
\begin{figure}
\begin{center}
\includegraphics[width=11.0cm]{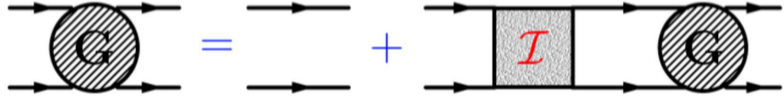}
\caption{Pictorial representation of the inhomogeneous integral equation
fulfilled by the four-leg Green's function. The first irreducible diagrams
contributing to the interaction kernel ${\cal I}$ are shown in Fig.
\ref{fig_kern}.
}\label{fig_GF4}
\end{center}
\end{figure}

\begin{figure}
\begin{center}
\includegraphics[width=9.cm]{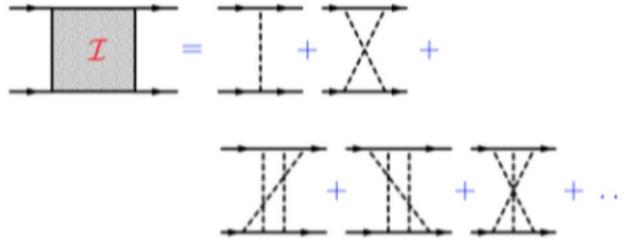}
\caption{ The lowest-order irreducible (see text) Feynman diagrams contributing to the
interaction kernel of the integral equation for the four-leg Green's
function, Eq. \eqref{G4}. In a full theory,  both  interaction vertexes and  propagators have to
be properly dressed.}\label{fig_kern}
\end{center}
\end{figure}

\begin{figure}
\begin{center}
\includegraphics [width=8.0cm] {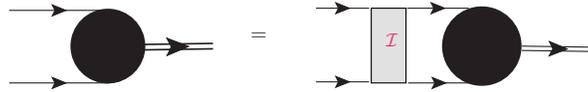}
\end{center}
\caption{Pictorial representation of the Bethe-Salpeter integral equation for a
bound system.} \label{fig_BSE}
\end{figure}

To briefly  illustrate  the main steps leading to the  BSE, let us consider a simple 
two-scalar case (without antiparticles).
The 4-point Green's Function ($\phi_i$ are the scalar fields) is given by
\be G(x_{1},x_{2};y_{1},y_{2})=<0\,|\,
T\{\phi_{1}(x_{1})\phi_{2}(x_{2})\phi_{1}^{+}(y_{1})
\phi_{2}^{+}(y_{2})\}\,|\,0>~~~,\ee 
and it fulfills the following  inhomogeneous integral equation, sketched in 
Fig. \ref{fig_GF4},
\be G=G_0~+~G_0~{\cal I}~G
\label{G4}\ee
where $G_0$ is the product of four free propagators and ${\cal I}$ is the interaction kernel.
It is 
given by the
infinite sum of {\em irreducible Feynman graphs}, i.e. the ones that cannot be
split into two lower order diagrams after cutting two  internal particle lines and without touching the 
 quantum lines \cite{BS51}. 
The set of the  irreducible diagrams up to the 6th power of the coupling constant 
are shown 
in  Fig. \ref{fig_kern}. The first diagram, though it
 contains two interaction vertexes (hence,  two   coupling
 constants), generates the infinite {\em ladder} series, after iterating the 
 integral
 equation in \eqref{G4}. The second diagram, that has four coupling
 constants, is the simplest crossed diagram. Finally, the last set 
 of diagrams contain crossed
 diagram with 6 interaction vertexes.  It is important to stress that each
 class of irreducible diagrams, 
 contributing to  the interaction kernel, in  turn
 gives rise to  an infinite series of diagrams by iteration. Such an observation
 allows us to
 anticipate that
 the solutions of the BSE corresponding to a truncated interaction kernel 
 ${\cal I}$ contains   however effects generated  by an infinite set of 
 powers of the coupling constant.
After properly inserting a complete 2-body Fock basis in 
$G(x_{1},x_{2};y_{1},y_{2})$, one can isolate
the bound state contribution
(assuming only one non-degenerate bound state, for the sake of simplicity)
that manifests itself as a pole, in momentum space, i.e.
\be  G(k,q;p)\Rightarrow G_B(k,q;p)\simeq {i\over (2\pi)^{-4}}~
\frac{{\chi(k;p_{B})~\bar{\chi}(q;p_{B})}}
{2\omega_{B}(p^{0}-\omega_{B}+i\epsilon)}
\label{bpole}\ee
where
 $\omega_B=\sqrt{M^2_B+|{\bf p}|^2}$,
 $p_B^\mu\equiv\{\omega_B,{\bf p}\}$ with $M_B$ the mass
of the bound state (the adopted metric is $\{1,-1,-1,-1\}$), $k$ is the relative four-momentum of the pair and 
 $\beta\equiv$ further quantum numbers. The function 
 $\chi(k;p_{B})$ is called the  BS
 amplitude, and 
 describes the residue together with its conjugate (notice that the conjugate
 $\bar \chi$ is actually defined through Eq. \eqref{bpole} and involves the
 chronological product, to be carefully considered \cite{Nakarev}). Unfortunately, the BS
 amplitude does not have a   probabilistic
interpretation, but one  can profitably retrieve a probabilistic environment once the LF framework and the Fock
expansion of the interacting two-body state is introduced (see Sect.
\ref{sec_lf}). In conclusion,
close to the bound-state pole, i.e. $p^0\to \omega_B=$
the Green's function is approximated by
\be G\simeq ~ {i\over (2\pi)^{-4}}~
\frac{{\chi(k;p_{B})~\bar{\chi}(q;p_{B})}}
{2\omega_{B}(p^{0}-\omega_{B}+i\epsilon)}~+~regular~terms
\label{G4pole}\ee
By going through  the following steps: i) insert the approximation
\eqref{G4pole} in both sides of
$ G=G_0+G_0 ~ {\cal I}~ G$, ii) multiply by $(p^0-\omega_B)$ and iii) look 
close to the pole,
one eventually gets
the BSE, viz (see Fig. \ref{fig_BSE})
\be   \chi(k;p_{B},\beta)= G_0(k;p_B,\beta) ~\int
d^{4}k'~ {\cal I}(k,k';p_B)~\chi(k';p_{B},\beta) \ee
The normalization of the BS amplitude is given by (see Ref. \cite{Lurie}, and
Refs. \cite{Nakarev,Nakanorm,Alkofer} for  further interesting issues on the BS
norm)
\be
\int {d^4 q\over (2\pi)^4} \int {d^4 k\over (2\pi)^4} ~
\bar \chi(q,p_B)
~ \left.{\partial \over \partial p_\mu} \Bigl[G^{-1}_0(k,p) (2\pi)^4 \delta^4(q-k)-i 
{\cal I}(q,k,p)\Bigr]\right|_{p^2=M^2_B}
\nonu
\times ~
\chi(k,p_B)=i2 p^\mu_B
\label{norm}
\ee
\section{ The Nakanishi  integral  representation  and
the BS Amplitude}
\label{sec_nir}
\begin{figure}
\begin{center}
\includegraphics[width=6.cm]{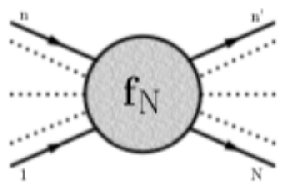}
\includegraphics[width=5.cm]{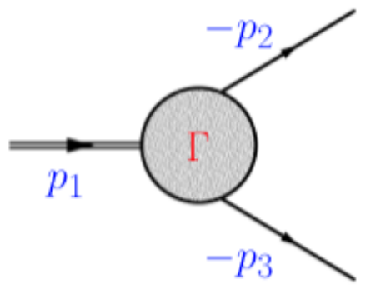}
\caption{Left panel: transition amplitude with $N$ external legs. The bubble represents the
infinite possible contributions, and the legs are represented just for
enumerating
 the external momenta, but they do not indicate any external
propagation.
Right panel: the 3-leg transition amplitude, relevant for the application to the
two-body Bethe-Salpeter amplitude (see text), the same previous caveat has to be
applied to the external legs.} \label{fig_TA}
\end{center}
\end{figure}

In the sixties, Noburo Nakanishi \cite{nak63,nak71} 
proposed an integral representation of
transition amplitudes (see Fig. \ref{fig_TA}), based on the parametric formula for the Feynman
diagrams.
Within the perturbation theory for scalars,    the transition
amplitude  with $N$ external legs, $f_N$, gets an infinite number of 
contributions,
whose
 generic expression is given by 
\be
f_{ \cal G}(p_1,p_2,...,p_N)\propto ~\prod_{i=1}^k\int d^4q_i~
\frac{1}{(\ell_1^2-m_1^2)(\ell_2^2-m_2^2)~\dots ~(\ell_n^2-m_n^2)}
\ee
where  $n$ indicates the total number of propagators,  $k$ is the total number of  loops
($\equiv$ number
of integration variables), $\ell_j$ are the internal momenta, combinations of 
 external
momenta $p_r$ and loop ones $q_i$.
The label 
${ \cal G}$ is a shorthand notation replacing $\{ n, k\}$.
By standard manipulations one can write
\be
f_{ \cal G}(\{p_i\})\propto ~\prod_{j=1}^n\int_0^1 d\alpha_j~
\frac{\delta(1-\sum_{r} \alpha_r)}
{U(\{\alpha_r\})~\left[\sum_{r} \alpha_r m^2_r
-\sum_{h} \eta_h(\{\alpha_i\})~ s_h +i\epsilon\right]^{n-2k}}
\label{Tacon}\ee
where  $\{\alpha_i\}$ are  $n$ Feynman parameters, 
$\{s_h\}$ all the {\em independent
scalar products} one can construct from the $N$ external legs with $1\le h\le
N'<N$,
$\eta_h(\{\alpha_r\})$ and 
 $U(\{\alpha_r\})$ are well-defined
combinations of the Feynman parameters \cite{nak71}.
 It is worth noticing that  the dependence upon $\{n,k\}$ affects 
  the denominator, and therefore each contribution to the total transition
 amplitude has its-own analytic structure.
Nakanishi introduced  a compact
and elegant expression of the full $N$-leg amplitude 
 $f_N(s)=\sum_{ \cal G} ~ f_{ \cal G}(s)$, by 
inserting in the generic contribution $f_{\cal G}$ the following identity
\be 
1\doteq \prod_{j=1}^{N'} \int_0^1dz_j~
\delta\left(z_j-\frac{\eta_j(\{\alpha_i\})}{\beta}\right )
\int_0^\infty d\gamma~\delta\left(\gamma-\sum_i\frac{\alpha_i m_i^2}{\beta}
\right)\ee
where 
 $\beta=\sum_h \eta_h(\{ \alpha_i\})$
  (see \cite{nak71} for details). After  integrating by parts $n-2k-1$ times
  the expression of $f_{\cal G}$, one gets from Eq. \eqref{Tacon}
\be f_{\cal G}(\tilde s)\propto~\prod_{j=1}^{N'}\int_0^1dz_j\int_0^\infty
d\gamma
\frac{\delta(1-\sum_j z_j)~\tilde{\phi}_{\cal G}(\vec z,\gamma)}
{{{(\gamma-\sum_j z_j~s_j)}}}\ee
 where $\tilde{\phi}_{\cal G}(\vec z,\gamma)$  is a proper weight function
  (that in this perturbative framework lives in the realm of the distribution
  functions),
with $\vec z\equiv\{z_1, z_2,~\dots~, z_{N'}\}$ and 
$\tilde s\equiv \{s_1,s_2,~\dots~, s_{N'}\} $.
Summarizing, the dependence upon the details of the diagram,
i.e. $\{n,k\}$, moves from
the denominator to the numerator.
Hence,  one has exactly the same formal
expression for  the denominator of any diagram ${\cal G}$.

The full  $N$-leg transition amplitude is the sum of infinite diagrams
${\cal G}$ and it can be formally written as
\be f_N(\tilde s)~=\sum_{\cal G} ~ f_{\cal G}(\tilde s)\propto~\prod_{j=1}^{N'}
\int_0^1dz_j
\int_0^\infty
d\gamma
\frac{\delta(1-\sum_j z_j)~\phi_N(\vec z,\gamma)}{(\gamma-\sum_i z_i~s_i)} \ee
where
$ {\phi}_N(\vec z,\gamma)=\sum_{\cal G}  \tilde{\phi}_{\cal G}(\vec z,\gamma) 
$
is called a Nakanishi weight function (NWF). It  is a real function, depending upon
$N'$ compact variables, $\vec z$, and one non compact, $\gamma$.

For the 3-leg transition amplitude, (cf the right panel in Fig. \ref{fig_TA})
after   eliminating one compact variable and  
 redefining
the NWF,  one reduces to
 \be f_3(p^2,k^2,k\cdot p)=\int_{-1}^1 dz
\int_{0}^\infty d\gamma \frac{\pi_3(z,\gamma)}
{\gamma+m^2-{p^2\over 4}- k^2-z k\cdot p-i\epsilon}
\label{vertex}\ee
with $p=p_1+p_2$ and $k=(p_1-p_2)/2$. Notice that $\phi_3$ is also known 
as the 
 vertex function, $\Gamma$, for
a scalar theory (N.B. for  fermions one has to add  spinor indexes).
 The expression holds  at {\em any}
  order in perturbation-theory.

Once we put one leg on the mass shell, and we assume $\phi_3(z,\gamma)$ 
as an unknown
functions, Eq. \eqref{vertex} becomes a natural choice as a trial function for
 obtaining actual
solution of BSE for a two-scalar interacting system
\cite{nak63,Kusaka,CK2006,FSV2}. The validation of this assumption, namely
translating an expression formally elaborated within a perturbative framework to a
non perturbative one, has been successfully obtained numerically,
 as discussed in what follows. This result should be not too much surprising if
 one takes into account the freedom associated to the NWF, that is the unknown
  to be determined.

A vertex function  $f_3(k,p)$  with one leg on mass-shell
is related to the
BS amplitude $\chi$ by attaching propagators to the two external virtual legs, i.e. 
schematically:
$ \chi = G_1\otimes G_2 \otimes f_3(\tilde s)$
where $G_i$ are the propagators of the two particles. Finally one writes
\be
\chi(k,p)=-i\int_{-1}^1 dz'
\int_{-\infty}^\infty d\gamma' \frac{g(z',\gamma';\kappa^2)}
{\left[\gamma'+\kappa^2- k^2-z' k\cdot p-i\epsilon\right]^3}
\ee
where $\kappa^2=m^2 -M^2/4$, with $m$ the mass of the constituent scalars and
$g(z',\gamma';\kappa^2)$ indicates a NWF corresponding to a given mass $M$ of the interacting system.
Indeed the lower extremum of $\gamma'$ has to be put equal to zero, in order to
avoid the spontaneous decay of the bound state.

\section{Projecting BSE onto the hyper-plane $x^+=0$ }
\label{sec_lf}

As mentioned above, NIR
yields the {\em needed freedom} for exploring non perturbative problems,
once the  NWFs are taken
as
 unknown real quantities.
 Unfortunately, even adopting   NIR, BSE still remains a  highly singular 
 integral
equation in the  Minkowski momentum space. The strategy adopted in Ref.
\cite{Kusaka} for obtaining actual solutions of the two-scalar homogeneous BSE,
in ladder approximation, relied on the uniqueness of the NWF. 
Profitably, a more general approach was introduced by Carbonell and Karmanov
\cite{CK2006,CK2006b}, that exploited the known analytic structure of the
BS amplitude, once NIR is applied.
Since the target  is to determine the NWF, one can perform 
analytic integrations, to formally obtain a new integral equation, more suitable
for the numerical treatment. The approach can be seen as a variant of the 3D
reduction, but in this case the assumed expression of the BS amplitude in terms
of NIR allows one the formally exact reconstruction of the full BS amplitude. In particular, in Ref.
\cite{CK2006} an explicitly-covariant LF framework \cite{CK_rev} was adopted,
while the non-explicitly covariant version can be found in Refs.
 \cite{FSV1,FSV2}. It
should be pointed out that the latter approach appears more suitable for isolating
and mathematically treating further 
 singularities one meets, particularly when the spin dofs are
involved (see Refs. \cite{dFSV1,dFSV2}). In general, the LF approach allows one a simpler treatment 
of the
analytic integration, since, e.g., double poles in the $k^0$ complex plane splits in pairs of single 
poles in the
$k^-$ and $k^+$ complex planes.

 In order to  obtain an integral equation for the NWF more tractable from
the numerical point of view   
(spin dofs lead to
   a {\em system} of  integral equations), there is an  illuminating step toward 
  the final goal. One can recognize 
  that within  the
LF framework a probabilistic interpretation is recovered by i) expanding
the BS amplitude on a LF Fock basis, and  ii) singling out the
{\em  valence component}. This is the amplitude of the Fock state with the lowest
number of constituents, and it has (together with all the other amplitudes in the
Fock expansion) the property to be  invariant under LF-boost transformations
(see Ref. \cite{Brod_rev}). In particular, the integral of the square modulus of the valence
component gives the probability to find only two  constituents in the
interacting state. In the case of the non-explicitly covariant LF framework,
the valence component is formally obtained by integrating the BS amplitude on
$k^-=k^0-k^3$.  Namely
\be {\rm Valence~ w.f.}=~\psi_{n=2}(\xi,k_\perp)=~{p^+\over \sqrt{2}}~\xi~(1-\xi)
\int {dk^- \over 2 \pi}
\chi(k,p)=
\nonu
={1\over \sqrt{2}}\xi~(1-\xi)~\int_0^{\infty}d\gamma'~
{g(\gamma',1-2\xi;\kappa^2) \over
[\gamma'+k_\perp^2 +\kappa^2+\left(2\xi-1\right)^2 {M^2\over 4}-i\epsilon]^2}
\ee
where $\xi=(1-z)/2$,   $M=2m-B$. with  $B$ the binding energy.
Notice that the above  mathematical step is  equivalent to make vanishing 
the relative  {\em LF-time  $x^+$} in the BS amplitude in coordinate space
(see, e.g.,\cite{FSV1,FSV2}), i.e. projecting the BS amplitude onto the hyperplane $x^+=0$. 
The above observations make attractive  to study what happens  when the 
 $k^-$ integration is applied to
  both sides of BSE. As a matter of fact,   on the lhs one    has (a part trivial factors) the valence component 
  that lives in a 3D
 space, and contains the NWF, while on the rhs one remains with a folding of the
 NWF with the so-called Nakanishi kernel, $ V^{LF}(\alpha;\gamma,z;\gamma',z')$, viz
\be
{\rm valence ~w.f.} ~\propto~ \int_0^{\infty}d\gamma'\frac{g(\gamma',z;\kappa^2)}{[\gamma'+\gamma
+z^2 m^2+(1-z^2)\kappa^2-i\epsilon]^2} =
\nonu=
\int_0^{\infty}d\gamma'\int_{-1}^{1}dz'~V^{LF}(\alpha;\gamma,z;\gamma',z')
~~g(\gamma',z';\kappa^2).
\label{nakafin}\ee
where  $\alpha$ is the coupling
constant. N.B.
 $ V^{LF}(\alpha;\gamma,z;\gamma',z')$,  is determined 
 by the irreducible kernel
${\cal I}(k,k',p)$ (see, e.g., \cite{FSV1,FSV2} for details). In presence of spin dofs, the  evaluation of the Nakanishi kernel, has been
carried out in ladder approximation, but it was shown that it is plagued by further singularities, as
recognized for the first time in  Ref. \cite{CK2010}. Fortunately, within the non-explicitly covariant
LF framework, those singularities can be isolated and mathematically integrated \cite{dFSV1,dFSV2} by
using the standard approach introduced in Ref. \cite{Yan}.  

In the numerical calculations of Refs \cite{FSV2,dFSV1,dFSV2,Diquark},  an orthonormal basis given 
by the Cartesian product of Laguerre and 
Gegenbauer polynomials (the first ones depends upon the non compact variable $\gamma$ and the second
ones upon the compact variable $z$) is adopted for expanding
 $g(\gamma,z;\kappa^2)$. In ladder approximation, once we assign the mass of the system, i.e. $\kappa^2$, 
  the integral equation in Eq. \eqref{nakafin} becomes
 a generalized
eigen-equation, with eigenvalue  $\alpha$, and the eigenvector
composed by the coefficients of
the expansion of $g(\gamma,z;\kappa^2)$.
If the eigen-equation  admits a solution, for a given mass $M$ (it is a non linear parameter) of the system,
then we know how to reconstruct the  BS
amplitude, and eventually we validate the  whole procedure from the 4D BSE to its projection onto the
$x^+=0$ hyperplane, or better its integration on the $k^-$ component of the relative four-momentum. 

\section{ Excerpt of the numerical results}
\label{sec_ris}
Many calculations have been performed for the two-scalar system by using the ladder approximation, i.e.
 a kernel with the following  massive scalar exchange
 $ {\cal K}_S={g^2/ [(k-k')^2-\mu^2+i\epsilon]}$,
 where the coupling $g$ is related to the coupling $\alpha $ in Eq. \eqref{nakafin} by
 $\alpha=g^2/(16\pi m^2)$. In particular a successful comparison between the results in 
 Ref. \cite{CK2006} (explicitly-covariant LF approach) and Ref. \cite{FSV2} (non-explicitly covariant
 LF framework) has been obtained  for eigenvalues and eigenvectors.
 Encouraged by this achievement,  both excited states \cite{Tomio2016} and even  scattering lengths
 \cite{FSV3} 
 have been studied.  There have been also
investigations of the cross-diagram contribution to the kernel (second diagram in Fig.
 \ref{fig_kern})   \cite{CK2006b}, and its effect on the calculation of the electromagnetic form factor
 \cite{gigante2017bound}.

 In the two-fermion case, the ladder kernel has been generalized to take into account i) the pseudo-scalar
 exchange and ii) the vector one (massless and massive). As mentioned above, in this case one has to solve
 a system of four integral equations for determining the four NWFs needed to describe an s-wave
 interacting system \cite{CK2010,dFSV1,dFSV2}. A part the fixing of the issue of the LF singularities generated by the presence
 of spin dofs, in Refs. \cite{dFSV1,dFSV2}
 an  important feature  has been pointed out: the theoretically expected behavior of the
  high-momentum tail of the valence component of a fermion-antifermion s-wave state, interacting through the
  exchange of a massless vector, has been recovered. This means that dynamical quantities can be addressed,
  teven with a simple interaction kernel.

\begin{figure}
\begin{center}
\includegraphics[width=9.0cm]{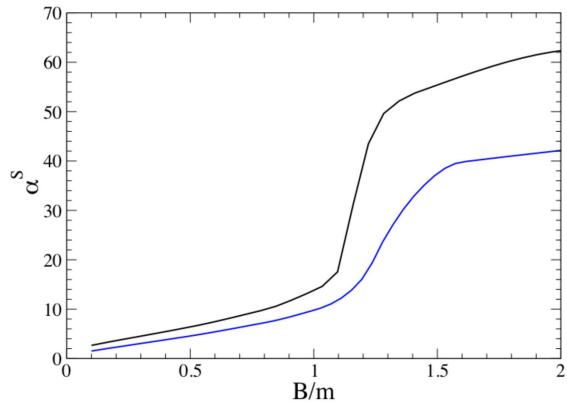}
\caption{ The coupling constant 
$\alpha=\lambda_F\lambda_S/(8\pi m)$, with $\lambda_{F(S)}$  the fermion (scalar) coupling in the
interaction Lagrangian, vs the binding energy per unit mass.  Blue solid line:  exchanged mass, 
$\mu/ m=0.15$.
Black solid line:  exchanged mass, $\mu/ m=0.5$}
\label{fig_FS}
\end{center}
\end{figure}

 The extensive collection of results and the agreement  achieved between different groups leads to  
 extend the application of NIR to other cases. In particular, some preliminary results for the 
 fermion-scalar system can be presented \cite{Diquark}. In this case,
the  BS amplitude  contains   two unknown scalar
functions $\phi_i$, viz
 \be
\Phi_{BS}(k,p)=\Bigl[\phi_1(k,p)+ {\psla k\over M}~\phi_2(k,p)\Bigl]~ U(p,s)
\ee
with
 $
(\psla p - M)~U(p,s)=0~
$.
 By applying NIR to each $\phi_i$, analogously to the two fermion case, one obtains a system of two
 coupled integral equations, and one can apply   the same tools for exactly
   transforming the initial ladder BSE in a generalized eigenvalue
 problem. In Fig. \ref{fig_FS}, preliminary results 
 for the coupling constants needed to bind a fermion-scalar system, with a given mass $M=2m-B$ 
 and interacting through a scalar exchange, are shown. The interesting fast growing that appears for
 $B/m\ge 1$ is a signature of the increasing relevance of the repulsion produced by the small component in 
 the Dirac spinor of
 the constituent fermion. The attraction of the scalar interaction is softened for increasing binding energy,
due to the competition between
 large and small components in $(\bar u ~u)$, at the fermionic vertex. Large $B/m$ values mean large kinetic energy,
 and in turn big effects from    
the small components.

.

\section{Conclusions \& Perspectives}
\label{sec_con}

 The    technique based on the Nakanishi integral representation of the BS amplitude 
  has to be considered a viable and effective tool  for solving 
BSE  Indeed,  the cross-check of
results obtained by different groups, for different interacting systems, with
kernels in ladder and cross-ladder contributions as well as with and without spin dofs,
 has produced a clear numerical evidence of the
validity of NIR for obtaining actual solutions. Noteworthy, the possibility to obtain the valence
component straightforwardly leads to  evaluate  relevant
quantities, like the the LF momentum distributions, so important for phenomenological investigations.of
bound systems in relativistic regime.
 In particular, one should 
  remind the second ingredient of the technique, i.e.  
the LF framework. It  has well-known advantages in performing
analytical
integrations, and in  the  fermionic  case it shows its effectiveness 
in full glory.

 The numerical validation of NIR strongly
encourages to face with the next challenges represented by
including self-energies and vertex corrections evaluated within the same
framework (works in progress on the Dyson-Schwinger Equations) as well as by  the
possible construction of interaction kernels beyond the sum of the first two contributions in Fig. \ref{fig_kern},
e.g., moving   
from the fully off-shell ladder series, 
that fulfills an integral equation,  to a closed form for the
 cross-ladder T-matrix.

%

%
%
\bibliographystyle{salme-spphys.bst}
\bibliography{salme-naka}
\end{document}